\documentstyle[preprint,aps]{revtex}
\newcommand{\eq}{\begin{equation}}
\newcommand{\en}{\end{equation}}
\newcommand{\eqa}{\begin{eqnarray}}
\newcommand{\ena}{\end{eqnarray}}
\newcommand{\eqs}{\begin{displaymath}}
\newcommand{\ens}{\end{displaymath}}
\newcommand{\eqas}{\begin{eqnarray*}}
\newcommand{\enas}{\end{eqnarray*}}

\begin{document}

\draft

\title{Soliton Solutions in Noncritical String Field Theory?
\footnote{Talk given by N.I. 
at ``Inauguration Conference of APCTP'', 4-10 June, 1996}}

\author{Nobuyuki Ishibashi and Hikaru Kawai}
\address{KEK Theory Group, 1-1 Oho Tsukuba Ibaraki 305, Japan}

\maketitle
\begin{abstract}
We look for soliton solutions in $c=0$ 
noncritical string field theory constructed 
by the authors and collaborators. It is shown that the string field action 
itself is very complicated in our formalism but it satisfies a very simple 
equation. We derive an equation which a solution to the equation of motion 
should satisfy. Using this equation, we conjecture the form of a soliton 
solution which is responsible for the nonperturbative effects of 
order  $e^{-A/\kappa}$. 
\end{abstract}


\section{Introduction}
Solitons in string theory play very important roles in understanding the 
dynamics of string theory\cite{HSP}. Especially D-branes\cite{Db}, 
giving a stringy description of solitons, may be very useful in exploring 
really stringy features of dynamics. In \cite{Dins}, the authors claim that 
D-instanton is relevant to nonperturbative effects in closed string theory. 
As was argued by Shenker \cite{shenker}, 
the strength of nonperturbative effects in 
closed string theory should be of order $e^{-A/\kappa}$, where $\kappa^2$ is 
the closed string coupling constant. 
Since the nonperturbative effects 
are of order $e^{-A/\kappa^2}$ in usual field theory,  
such nonperturbative effects are considered to be stringy. 
In \cite{Dins}, the contribution of D-instanton to 
the partition function was shown to be of the form $e^{-A/\kappa}$, where 
$A/\kappa$ here corresponds to the disk amplitude with the Dirichlet boundary 
condition. Hence we may be able to attribute the nonperturbative effects of 
closed string theory to D-instantons. Similar ideas using other solitons 
were proposed in \cite{WBBS}. 

If these ideas are true, it will be a very important development in string 
theory, leading to a quantitative nonperturbative treatment of string theory. 
Exactly solvable noncritical string theory\cite{DS} will be a good place to test 
such ideas. Nonperturbative effects in noncritical string theory were shown to 
have the form $e^{-A/\kappa}$. We expect that there exist soliton solutions 
whose contribution to the partition function is of  order $e^{-A/\kappa}$. 
It is very intriguing to find and study such solutions and see if the above 
claim about the role of D-instantons in string theory is true or not. 
What we would like to do in this paper is to look for such solutions. 

In order to do so, we need a string field theory.  
We need to know what is the equation of motion for the string field, which 
should be solved to obtain the soliton solution. We need to know what is the 
string field action to which the value $A/\kappa$ should correspond. 
In the references in \cite{david}, this problem was considered in the matrix 
model framework or the $c=1$ string field theory written in terms of the 
eigenvalue distribution of the matrices. 
It was claimed that the nonperturbative effect of the form $e^{-A/\kappa}$ is 
due to the eigenvalue tunneling and the soliton solution corresponding to 
such an effect was identified. However it is difficult to see what kind 
of soliton it is in the string theory point of view. Since the eigenvalue 
distribution is not the quantity we usually use in critical string theory, 
the interpretation of the tunneling of eigenvalues seems cumbersome. 

In this paper we will use the noncritical string field theory 
constructed by the authors and collaborators\cite{IK} to consider this problem. 
The advantage of this formalism is that 
the string field is a functional of the configuration of 
string as in the usual string field theories. 
We will look for soliton solutions which is responsible for the nonperturbative 
effect of the form $e^{-A/\kappa}$. 

The organization of this paper is as follows. In section 2 and section 3, we 
will review our string field theory. In section 2, we present the stochastic 
quantization of usual field theory in a form which is convenient for considering 
string field. 
In section 3, we present our string field theory as a 
generalization of stochastic quantization. Throughout this paper, we restrict 
ourselves to $c=0$ example, although many of the results hold for $c\neq 0$ 
without any modifications. 
In section 4, we consider string 
field action in our string field theory. We obtain a simple functional 
differential equation which 
this string field action satisfies. 
It turns out that the string field 
action itself, as a solution of this functional differential equation, 
is very complicated. In section 5, we will show that a solution of equation of 
motion should satisfy a simple equation. Although this equation is not enough 
for obtaining the soliton solution we are looking for, we can conjecture the 
form of the soliton solution. Section 6 is devoted to discussions. 

\section{Stochastic Quantization}
Stochastic quantization is a way of quantization invented by Parisi and 
Wu\cite{PW}. Here we present it in a form modified from its original form 
for later convenience. More detailed discussions including the relation 
between the presentation here and the Parisi-Wu's formulation is given in
\cite{Proc}.

Let us consider a $D$-dimensional Euclidian scalar field theory with 
action $S[\phi ]$ as an example. 
The quantities of interest are the correlation functions:
\eq
<\phi (x_1)\cdots \phi (x_n)>=
\frac{\int [d\phi ]e^{-S[\phi ]}\phi (x_1)\cdots \phi (x_n)}
{\int [d\phi ]e^{-S[\phi ]}}.
\en

In stochastic quantization such quantities are calculated as follows. 

\begin{enumerate}
\item  First we set up a fictitious operator system. 
Corresponding to the field $\phi (x)$, 
we consider the creation and the annihilation operators 
$\hat{\phi}^\dagger (x), \hat{\phi}(x)$ satisfying 
\eq
[\hat{\phi}(x), \hat{\phi}^\dagger (y)]=\delta^D(x-y).
\en
Notice that this is not the usual equal time commutation relation. 
These operators act on states generated from the bra and ket vacua $|0\rangle ,
~\langle 0|$ 
satisfying
\eqa
& &
\hat{\phi}(x)|0\rangle =\langle 0|\hat{\phi}^\dagger (x)=0,
\nonumber
\\
& &
\langle 0|0\rangle =1.
\ena

\item In this system, we consider the fictitious time $t$ and the time 
evolution operator $\hat{H}$ which has the following form:
\eq
\hat{H}=\frac{1}{2}\int d^Dx[\frac{\delta S}{\delta \phi (x)}
[\hat{\phi}^\dagger (x)]\hat{\phi}(x)-\hat{\phi}^2(x)].
\label{pham}
\en
 
\item Then the correlation function can be expressed as 
\eq
<\phi (x_1)\cdots \phi (x_n)>=
\lim_{t\rightarrow \infty}
\langle 0|e^{-t\hat{H}}\hat{\phi}^\dagger (x_1)\cdots \hat{\phi}^\dagger (x_n)
|0\rangle .
\label{corr}
\en
\end{enumerate}

Eq.(\ref{corr}) can be proven as follows. Let us define the probability 
distribution functional $P[\phi ;t]$:
\eq
P[\phi ;t]\equiv
\langle 0|e^{-t\hat{H}}\prod_x\delta(\hat{\phi}^\dagger (x)-\phi (x))
|0\rangle .
\label{PDF}
\en
Then eq.(\ref{corr}) means 
\eq
<\phi (x_1)\cdots \phi (x_n)>=
\int [d\phi ]\lim_{t\rightarrow \infty}P[\phi ;t]
\phi (x_1)\cdots \phi (x_n).
\label{corr1}
\en
In order to obtain $\lim_{t\rightarrow \infty}P[\phi ;t]$, one should notice 
that $P[\phi ;t]$ satisfies the 
Fokker-Planck equation:
\eqa
\partial_tP(t;\phi (x))
&=&
-\langle 0|e^{-t\hat{H}}\hat{H}\prod_x\delta(\hat{\phi}^\dagger (x)-\phi (x))
|0\rangle 
\nonumber
\\
&=&
\frac{1}{2}\int dx^D(\frac{\delta^2}{\delta \phi (x)\delta \phi (x)}
+\frac{\delta}{\delta \phi(x)}\frac{\delta S}{\delta \phi (x)})
P[\phi ;t].
\label{FP}
\ena

In the limit $t\rightarrow \infty$, we expect that $P[\phi ;t]$ goes to 
the stationary distribution, i.e. 
\eq
\frac{1}{2}\int dx^D\frac{\delta}{\delta \phi (x)}
(\frac{\delta}{\delta \phi (x)}
+\frac{\delta S}{\delta \phi (x)})
\lim_{t\rightarrow \infty}P[\phi ;t]=0. 
\en
This equation has one obvious solution which is 
\eq
\lim_{t\rightarrow \infty}P[\phi ;t]=Ce^{-S[\phi ]}.
\label{sol}
\en
The constant $C$ can be fixed as $C=(\int [d\phi ]e^{-S[\phi ]})^{-1}$, because 
eq.(\ref{PDF}) implies $\int [d\phi ]P[\phi ;t]=1$. This distribution 
is exactly what is needed to show eq.(\ref{corr1}). 
Actually if the action $S$ satisfies some general conditions, one can prove 
that $P[\phi ;t]$ indeed goes to the stationary distribution in eq.(\ref{sol}) 
in the limit $t\rightarrow \infty$. 

\section{``Stochastic Quantization'' of String Theory}
As we saw in section 2, stochastic quantization is a way to rewrite field 
theory. In this section, we will see that a generalization of it is quite 
useful in noncritical string field theory. 

Here let us consider $c=0$ noncritical string theory as an example. The string 
field is a functional of the configuration of string. In $c=0$ case, only 
degrees of freedom on the worldsheet is the metric. Thus the only invariant 
quantity which specifies the configuration of a string  is its length $l$. 
Therefore the string field $\Psi (l)$ is a function of length. 
The quantities of interest are the correlation functions such as  
$<\Psi (l_1)\cdots \Psi (l_n)>$, which corresponds to the worldsheet with 
$n$ boundaries of lengths $l_1,\cdots ,l_n$. These are the quantities usually 
called loop amplitudes. 
The ``stochastic quantization'' of this system of string field goes as follows. 

\begin{enumerate}
\item Corresponding to the field $\Psi (l)$, 
we consider the creation and the annihilation operators 
$\hat{\Psi}^\dagger (l), \hat{\Psi}(l)$ satisfying 
\eq
[\hat{\Psi}(l), \hat{\Psi}^\dagger (l')]=\delta (l-l').
\en
These operators act on states generated from the bra and ket vacua $|0\rangle 
,~\langle 0|$ 
satisfying
\eqa
& &
\hat{\Psi}(x)|0\rangle =\langle 0|\hat{\Psi}^\dagger (x)=0,
\nonumber
\\
& &
\langle 0|0\rangle =1.
\ena

\item In this system, we consider the fictitious time $t$\footnote{This time 
variable was first introduced on the string worldsheet in \cite{KKMW}.}
 and the time 
evolution operator $\hat{H}$\cite{IK} which has the following form:
\eqa
\hat{H}
&=&
\int_0^{\infty}dl_1\int_0^{\infty}dl_2
  \hat{\Psi}^{\dagger}(l_1)\hat{\Psi}^{\dagger}(l_2)\hat{\Psi} (l_1+l_2)(l_1+l_2)
\nonumber
\\
& &
  +\kappa^2\int_0^{\infty}dl_1\int_0^{\infty}dl_2
  \hat{\Psi}^{\dagger}(l_1+l_2)\hat{\Psi} (l_1)\hat{\Psi} (l_2)l_1l_2
\nonumber
\\
& &
+\int _0^{\infty}dl(3\delta ''(l)-\frac{3}{4}\lambda\delta (l))\hat{\Psi} (l).
\label{sham}
\ena
Here $\kappa^2$ is the coupling constant of string and $\lambda$ is the 
cosmological constant on the worldsheet. 
 
\item Then the correlation function can be expressed as 
\eq
<\Psi (l_1)\cdots \Psi (l_n)>=
\lim_{t\rightarrow \infty}
\langle 0|e^{-t\hat{H}}\hat{\Psi}^\dagger (l_1)\cdots \hat{\Psi}^\dagger (l_n)
|0\rangle .
\label{scorr}
\en
\end{enumerate}

Eq.(\ref{scorr}) was proven in \cite{wata}\cite{JR}. The procedure here 
is almost the same as that in section 2, but there is one crucial difference. 
In section 2, we start from a system with action $S$ and construct the 
time evolution operator $\hat{H}$ from $S$. 
Here we construct $\hat{H}$ directly without any 
informations of $S$. Although the time evolution operator $\hat{H}$ 
here consists of 
simple processes of splitting and joining of strings, it is not of the form 
as in eq.(\ref{pham}). Therefore at least in terms of the variable $\Psi (l)$, 
it is impossible to interpret that this $\hat{H}$ 
comes from some $S$ following the 
procedure in section 2.\footnote{The result in \cite{JR} suggests that 
by changing the choice of variable, such an interpretation may be possible.} 

Since $\hat{H}$ is not constructed from an action, 
it is impossible to prove that 
this system converge towards equilibrium in the limit $t\rightarrow\infty$ from 
some properties of the action. However from the matrix model results, one can 
prove that it is indeed the case and 
\eq
\lim_{t\rightarrow\infty}
\langle 0|e^{-t\hat{H}}\hat{H}=0.
\en
Actually one can show a stronger equation
\eq
\lim_{t\rightarrow\infty}
\langle 0|e^{-t\hat{H}}l\hat{T}(l)=0,
\label{vir}
\en
where 
\eq
l\hat{T}(l)=
l\int_0^{\infty}dl'
  \hat{\Psi}^{\dagger}(l')\hat{\Psi}^{\dagger}(l-l')
+\kappa^2l\int_0^{\infty}dl'
  \hat{\Psi}^{\dagger}(l+l')\hat{\Psi} (l')l'
+3\delta ''(l)-\frac{3}{4}\lambda\delta (l),
\en
and
\eq
\hat{H}=\int_0^{\infty}dll\hat{T}(l)\hat{\Psi}(l).
\en
Indeed eq.(\ref{vir}) generates \cite{IK} the Schwinger-Dyson equations 
which are equivalent to the Virasoro constraints \cite{FKN}.

\section{String Field Action}
In section 3, we obtain a simple time evolution operator $\hat{H}$ consisting of 
up to three string vertices. However what we need is a string field action 
in order to consider soliton solutions in string theory. Although $\hat{H}$ is 
not in such a form that the action $S$ can simply be read off as in section 
2 case, it is possible to construct it in a brute force manner. 

Following the procedure in section 2, let us define the probability distribution 
functional $P[\Psi ;t]$:
\eq
P[\Psi ;t]\equiv
\langle 0|e^{-t\hat{H}}\prod_l\delta(\hat{\Psi}^\dagger (l)-\Psi (l))
|0\rangle .
\label{sPDF}
\en
As in section 2, it satisfies the Fokker-Planck equation, 
\eq
\partial_t P[\Psi ;t]=\int dl
\frac{\delta}{\delta \Psi (l)}
[l\int_0^ldl'\Psi (l')\Psi (l-l')
-\kappa^2l\int_0^\infty dl'l'\Psi (l+l')\frac{\delta}{\delta \Psi (l')}
+3\delta ''(l)-\frac{3}{4}\lambda\delta (l)]P[\Psi ;t].
\label{sFP}
\en

Since 
\eq
<\Psi (l_1)\cdots \Psi (l_n)>=
\int [d\Psi ]\lim_{t\rightarrow \infty}P[\Psi ;t]
\hat{\Psi} (l_1)\cdots \hat{\Psi} (l_n),
\label{scorr1}
\en
$\lim_{t\rightarrow \infty}P[\Psi ;t]$ should be expressed as 
\eq
\lim_{t\rightarrow \infty}P[\Psi ;t]=\frac{e^{-S/\kappa^2}}
{\int [d\Psi ]e^{-S/\kappa^2}},
\en
by the string field action $S[\Psi ]$. From the fact stated in the last 
paragraph of section 3, in the limit $t\rightarrow \infty$, 
$P[\phi ;t]$ converges toward the static equilibrium 
distribution satisfying 
\eq
[l\int_0^ldl'\Psi (l')\Psi (l-l')
-\kappa^2l\int_0^\infty dl'l'\Psi (l+l')\frac{\delta}{\delta \Psi (l')}
+3\delta ''(l)-\frac{3}{4}\lambda\delta (l)]
\lim_{t\rightarrow \infty}P[\Psi ;t]=0.
\en
Therefore the string field action satisfies the following equation:
\eq
l\int_0^ldl'\Psi (l')\Psi (l-l')
+l\int_0^\infty dl'l'\Psi (l+l')\frac{\delta S}{\delta \Psi (l')}
+3\delta ''(l)-\frac{3}{4}\lambda\delta (l)=0.
\label{Seq}
\en

Eq.(\ref{Seq}) is an equation which should be solved to express $S$ as 
a functional of $\Psi$. It is possible to obtain a solution to 
eq.(\ref{Seq}) perturbatively as follows. 
As an equation for $\Psi$, it is in the same form as the Schwinger-Dyson 
equation for the disk amplitude in the backgroud 
$J(l)=\frac{\delta S}{\delta \Psi (l)}$ \cite{IK}. 
Therefore it has a solution
\eq
\Psi (l)=
\sum_n\frac{1}{n!}
\int_0^\infty dl_1\cdots \int_0^\infty dl_n
<\Psi (l)\Psi (l_1)\cdots\Psi (l_n)>_0
J(l_1)\cdots J(l_n).
\label{psi}
\en
Here $<\Psi (l_1)\cdots\Psi (l_n)>_0$ denotes the genus $0$ loop amplitude. 
By solving eq.(\ref{psi}) iteratively, one can express $J(l)$ in terms of 
$\Psi (l)$ and eventually a solution $S_0[\Psi ]$ to eq.(\ref{Seq}) 
can be obtained as an expansion in terms of $\Psi -<\Psi >_0$:
\eq
S_0[\Psi ]=\frac{1}{2}(\Psi -<\Psi >_0){\cal K}^{-1}(\Psi -<\Psi >_0)+
{\cal O}((\Psi -<\Psi >_0)^3).
\label{saction}
\en
${\cal K}$ is an operator which acts on a function $f(l)$ as  
$({\cal K}f)(l)=\int_0^\infty dl'<\Psi (l)\Psi (l')>_0f(l')$.
It is easily seen that the tree level calculations using this action 
reproduce the tree level amplitudes $<\Psi (l_1)\cdots\Psi (l_n)>_0$. 
The first few terms in the expansion of such an action 
were calculated in \cite{MS}. 
Since there is no reason for this expansion to terminate, it seems that this 
string field action is nonpolynomial. 

We have not checked if  $S_0[\Psi ]$ is the actual string field action which 
reproduces all the higher loop amplitudes. 
$S_0[\Psi ]$ may not be the only solution to eq.(\ref{Seq}). 
If  $S_0[\Psi ]$ fails to reproduce one-loop amplitudes, 
we should add a correction term as $S_0[\Psi ]+\kappa^2S_1[\Psi ]$ to make it 
do so. Repeating the same procedure for higher loop amplitudes, the string 
field action $S[\Psi ]$ may be able to be obtained as 
\eq
S[\Psi ]=S_0[\Psi ]+\kappa^2S_1[\Psi ]+\kappa^4S_2[\Psi ]+\cdots .
\label{corrections}
\en
The structure of the correction terms $S_n[\Psi ]$ will be severely restricted 
by the fact that $S[\Psi ]$ is a solution to eq.(\ref{Seq}). 

\section{Soliton Solutions?}
Thus our string field action seems to be a nonpolynomial action which 
may include 
infinitely many terms needed for reproducing higher loop amplitudes. 
Such properties are exactly what  
we expect for the action of a covariant string field theory of critical 
string. It may look impossible to solve the classical equation of motion 
for such a nonpolynomial action. 

However, since the string action $S$ satisfies eq.(\ref{Seq}), one can obtain 
informations about solutions to the equation of motion. 
If $\Psi (l)$ is a solution to the equation of motion
\eq
\frac{\delta S}{\delta \Psi (l)}=0,
\en
it should satisfy 
\eq
l\int_0^ldl'\Psi (l')\Psi (l-l')
+3\delta ''(l)-\frac{3}{4}\lambda\delta (l)=0.
\label{eqm}
\en
This equation can be solved easily once Laplace transformed:
\eq
\partial_\zeta (\tilde{\Psi}(\zeta ))^2-3\zeta^2+\frac{3}{4}\lambda =0.
\en
Here $\tilde{\Psi}(\zeta )=\int dle^{-\zeta l}\Psi (l)$. The solution is 
of the form
\eq
\tilde{\Psi}(\zeta )=\pm \sqrt{\zeta^3-\frac{3}{4}\lambda\zeta +C},
\label{seqm}
\en
where $C$ is an integration constant. 

What we have proved is that if there is a solution to the equation of motion, 
it should be of the form in eq.(\ref{seqm}). It is not clear what choice of 
$C$ actually corresponds to a solution. One obvious choice is 
\eq
\tilde{\Psi}(\zeta )=
\sqrt{\zeta^3-\frac{3}{4}\lambda\zeta +\frac{\lambda^{3/2}}{4}}
=(\zeta -\frac{\sqrt{\lambda}}{2})\sqrt{\zeta +\sqrt{\lambda}},
\en
which coincides with the disk amplitude $<\Psi >_0$. 
This solution is special so that $\lim_{l\rightarrow\infty}\Psi (l)=0$ and 
$\Psi (l)>0$. Because of these nice properties the perturbative expansion 
around this solution is well-defined. 

For other choices of $C$, the solution does not have such properties and 
the perturbative expansion is ill-defined. The large $t$ limit in section 
3 does not seem to exist perturbatively. However, 
we can not discard such solutions for this reason, 
because the perturbation series is expected to become 
unstable around the soliton we are looking for \cite{shenker}\cite{david}. 

Hence, 
without the explicit form of the action $S[\Psi ]$, it is impossible to 
see which $C$ should be taken for the soliton solution, but 
here we conjecture that the following form of solution corresponds to 
the soliton solution:
\eq
\tilde{\Psi}(\zeta )=
\sqrt{\zeta^3-\frac{3}{4}\lambda\zeta +\frac{\lambda^{3/2}}{4}+
f(\kappa ,\lambda )}.
\label{soliton}
\en
Here $f(\kappa ,\lambda )$ is a function of $\kappa$ and $\lambda$ and 
$f(\kappa ,\lambda )\sim 0$ when $\kappa\sim 0$. Notice that the appearance 
of such a function depending on $\kappa$ is possible when the correction 
terms in eq.(\ref{corrections}) are not zero. 

The reason for this conjecture is the following. 
When $\kappa $ is small, this solution can be expanded as
\eq
\tilde{\Psi}(\zeta )\sim 
(\zeta -\frac{\sqrt{\lambda}}{2})\sqrt{\zeta +\sqrt{\lambda}}
+\frac{\frac{1}{2}f(\kappa ,\lambda )}
{(\zeta -\frac{\sqrt{\lambda}}{2})\sqrt{\zeta +\sqrt{\lambda}}}.
\en
This function has a pole 
at $\zeta =\frac{\sqrt{\lambda}}{2}$ and a cut 
for $\zeta \leq -\sqrt{\lambda}$. 
The pole here corresponds to a very short cut near 
$\zeta =\frac{\sqrt{\lambda}}{2}$ in eq.(\ref{soliton}). 
Since $\Psi$ corresponds to a disk amplitude, 
the imaginary part of it gives the distribution function of 
the matrix eigenvalues of the matrix model in the continuum limit. 
Therefore the eigenvalues are distributed in the region 
$\zeta \leq -\sqrt{\lambda}$ and at the point $\zeta =\frac{\sqrt{\lambda}}{2}$. 
This is exactly the eigenvalue distribution which people considered in 
\cite{david} to explain the form of the nonperturbative effect $e^{-A/\kappa }$. 
Therefore if the eigenvalue tunneling discussed in \cite{david} is responsible 
for such a nonperturbative effect, the soliton we should look for in our 
formalism is of this form.

\section{Discussions}
In this paper we look for soliton solutions in our string field theory 
of noncritical string. In order to do so, the equation which the string 
field action satisfies is very useful. We have shown that the string field 
action may be very complicated although the time evolution operator $\hat{H}$ in 
the stochastic quantization formulation is very simple. Perhaps such a 
situation occurs also for the critical string case. The covariant action 
is known to be very complicated but there can exist a very simple 
formulation using stochastic quantization. 

As for the soliton solution, we conjecture that it will be of the form 
in eq.(\ref{soliton}). Unfortunately we cannot fix the form of the function $f$. 
Without doing so, it is impossible to answer the questions like what kind of 
soliton it is etc.. 
To pursue further, we might need to look into the form of the action $S$ 
more carefully. 

\section*{Acknowledgements}
\hspace{5mm}
We would like to thank Sumit R. Das for useful discussions. 
N.I. would like to thank the 
organizers of the conference for giving him a chance to give a talk at 
this historical conference.

\end{document}